\documentclass[conference]{IEEEtran}

\IEEEoverridecommandlockouts

\usepackage{cite}      
\usepackage{amssymb,amsfonts}
\usepackage[cmex10]{amsmath}
\usepackage{algorithmic}
\usepackage{algorithm}
\usepackage{textcomp}
\usepackage{xcolor}
\usepackage{bm} 
\usepackage{amsthm}

\usepackage{booktabs}  

\usepackage{cases}
\usepackage{bbm}
\usepackage{subfigure}
\usepackage{balance}
\usepackage{stfloats}
\usepackage{graphicx,float}
\usepackage{epstopdf}
\usepackage{float}
\usepackage{tabularx}

\usepackage{verbatim}
\usepackage{url}

\begin{document}

\title{Hardware Impairment Parameter Estimation in Fluid Antenna Systems by Inverse Optimization 
	\thanks{\copyright~2026 IEEE. Personal use of this material is permitted.
		Permission from IEEE must be obtained for all other uses, in any current
		or future media, including reprinting/republishing this material for
		advertising or promotional purposes, creating new collective works,
		for resale or redistribution to servers or lists, or reuse of any
		copyrighted component of this work in other works.}}

\author{
	\IEEEauthorblockN{Songling Zhang\IEEEauthorrefmark{1},
		Zhaohui Yang\IEEEauthorrefmark{1},
		Ruopeng Xu\IEEEauthorrefmark{1},
	    Hao Xu\IEEEauthorrefmark{2},
        Xusheng Zhu\IEEEauthorrefmark{3},
        Zhaoyang Zhang\IEEEauthorrefmark{1},
	    and Kai-Kit Wong\IEEEauthorrefmark{3}
}

\IEEEauthorblockA{\IEEEauthorrefmark{1}College of Information Science and Electronic Engineering, Zhejiang University, Hangzhou, China}
\IEEEauthorblockA{\IEEEauthorrefmark{2}School of Information Science and Engineering,
Southeast University, Nanjing, China}
\IEEEauthorblockA{\IEEEauthorrefmark{3}Department of Electronic and Electrical Engineering,
University College London, London, UK}
\IEEEauthorblockA{Emails: sl-zhang@zju.edu.cn, yang\_zhaohui@zju.edu.cn, ruopengxu@zju.edu.cn, hao.xu@seu.edu.cn, \\ xusheng.zhu@ucl.ac.uk, ning\_ming@zju.edu.cn, kai-kit.wong@ucl.ac.uk}

}

\maketitle

\begin{abstract}

In this paper, the problem of estimating the hardware impairment parameters of fluid antennas (FAs) in a multi-base station (BS) multi-user communication system is studied. The actual hardware coefficients are unknown, whereas the FA positions and transmit beamforming vectors adopted by the BSs are known. We first formulate a forward spectral efficiency (SE) maximization problem and introduce aggregate interference variables whose consistency with the physical inter-cell interference can be maintained through a pricing mechanism. For inverse estimation, the aggregate interference is obtained from receiver-side interference measurements associated with the observed network decisions. We then construct a trust-region linear model of the position-update problem around the observed FA positions. To extend the applicability of the proposed framework to antenna positions obtained from inexact or heuristic algorithms, we develop an inverse optimization model with an explicit duality-gap tolerance, which is solved using sequential quadratic programming (SQP).
Simulation results show that the proposed scheme attains a lower  estimation error than KKT residual minimization (KKT-RM) at both considered transmit-power levels and improves the  SE by up to 22.86\% over the Ideal-hardware-assumption scheme.
\end{abstract}


\section{Introduction}

Fluid antennas (FAs) have emerged as a pivotal enabling technology for next-generation wireless networks, owing to the ability to flexibly adjust antenna positions \cite{9650760, 10753482}. By enabling  position adjustment within a given spatial region, fluid antenna systems (FASs) can achieve significant diversity  gain, outperforming traditional fixed-position antenna arrays \cite{11098630, 10130117, xu2023capacity}. However, the practical deployment of FAS in base stations (BSs) and user equipments (UEs) faces challenges due to hardware imperfections and dielectric losses.  FAS components are typically exposed to harsh outdoor environments, making them susceptible to long-term wear and tear caused by factors such as solar radiation, extreme temperature fluctuations, and humidity \cite{8581475}. These environmental stressors inevitably lead to hardware impairments, which manifest as unknown hardware impairment  coefficients that degrade the antenna's radiation efficiency \cite{11224643}. Consequently, optimizing network performance based on ideal, factory-calibrated antenna models leads to substantial inaccuracies and suboptimal resource allocation. Therefore, accurate estimation of the actual and in-situ  hardware impairment  coefficients of the FAS is paramount for precise network modeling and optimization. To tackle this issue, inverse optimization can be employed as a promising technique, which facilitates the recovery of unknown hardware parameters by mathematically inverting the forward model leveraging the observed  decision variables.

A number of works have investigated the applications of inverse optimization across diverse domains \cite{chan2020inverse,9336679,chan2025inverse,ahuja2001inverse,mohajerin2018data,7859377}. In particular, the work in \cite{chan2020inverse} studied inverse optimization methodologies for imputing unspecified left-hand-side constraint coefficients and uncertainty set parameters.  In \cite{9336679}, the authors developed an inverse optimization framework to learn a control policy that maps input states to optimal actions.  In \cite{chan2025inverse}, the authors provided a comprehensive review of inverse optimization, establishing a systematic framework that distinguishes between classical models and data-driven approaches. The work in \cite{ahuja2001inverse} developed a unified framework for inverse linear programming and network optimization problems. The work in \cite{mohajerin2018data} proposed a data-driven distributionally robust inverse optimization framework. The authors in \cite{7859377} proposed a data-driven inverse optimization framework to forecast the short-term aggregate demand of price-responsive loads.
However, most of these existing studies \cite{chan2020inverse,9336679,chan2025inverse,ahuja2001inverse,mohajerin2018data,7859377} focus on applications in automation, transportation, or economics, with limited exploration in the domain of wireless communications. Specifically, the application of inverse optimization to estimate physical hardware impairments in  FAS remains largely unexplored. Considering the significant impact of hardware impairment coefficients on FAS performance, bridging this gap is essential for the practical deployment and optimization of  wireless networks.

The main contribution of this paper is to develop an inverse optimization framework for estimating the unknown hardware impairment coefficients of FAs. We consider a multi-cell system in which the FA positions and beamforming vectors adopted by the BSs are known, while receiver-side interference measurements are associated with these known decisions. An aggregate-interference formulation is used to clarify how consistency pricing decouples the forward problem without treating interference as a freely adjustable variable. Rather than imposing a closed-form beamformer, we condition the position-update model on the known beamforming matrices and retain both intra-cell interference and inter-cell externalities. We subsequently construct a trust-region linear approximation of this model around the observed positions. Finally, a duality-gap relaxation of the inverse linear model is developed to accommodate positions generated by inexact or heuristic forward algorithms, and an sequential quadratic programming (SQP) algorithm is employed to estimate the hardware coefficients.
Simulation results show that the proposed scheme yields a lower  estimation error than KKT-RM at both 20 and 30 dBm and provides up to 22.86\%  SE improvement over the Ideal-hardware-assumption scheme.

\section{System Model and Forward Problem Formulation}

\subsection{System Model}
\begin{figure}[t]
	\centering
	\includegraphics[width=0.95\linewidth]{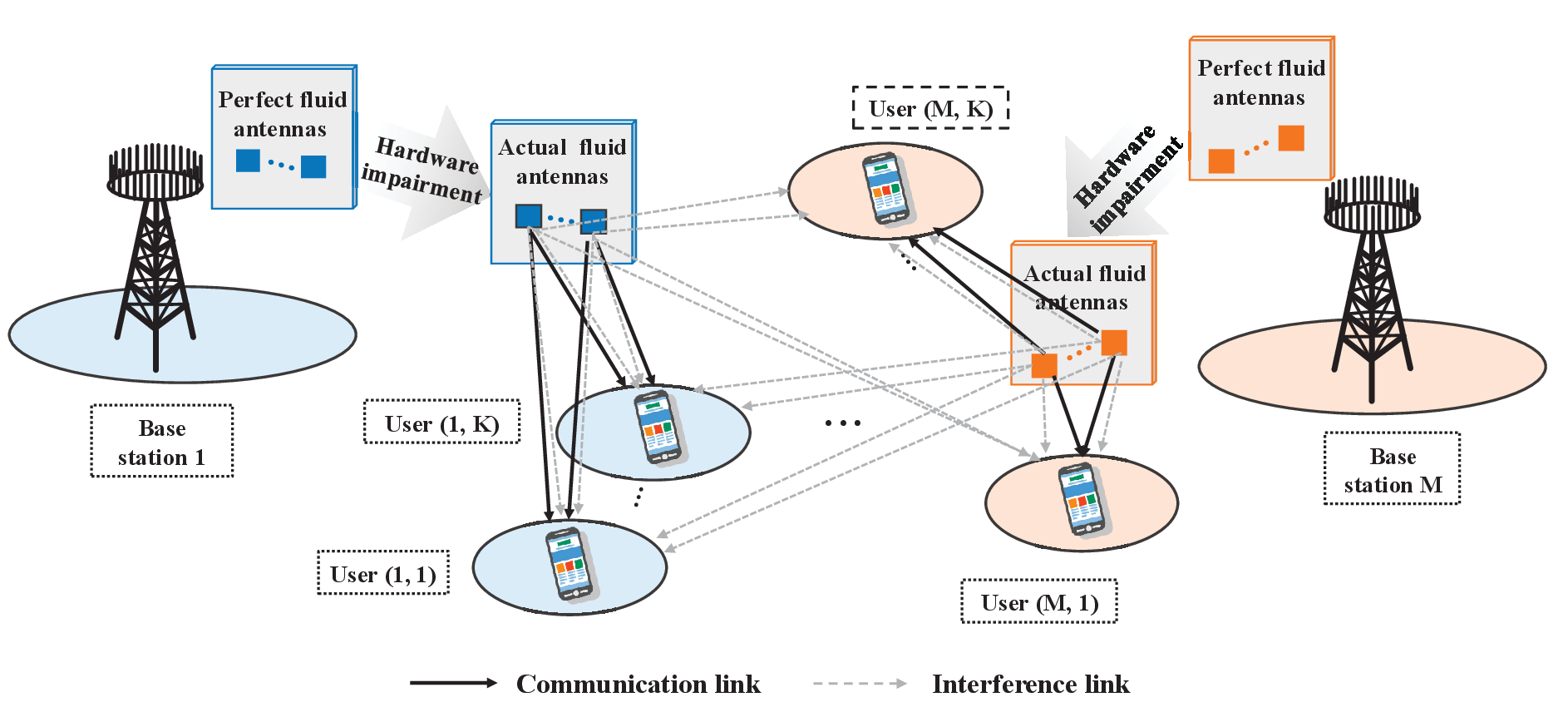}
	\caption{ The considered multi-BS multi-user 
communication system.}\label{fig 1}
\vspace{-0.2cm}
\end{figure}

As shown in Fig. \ref{fig 1}, we consider a downlink multi-BS multi-user  communication system consisting of $B$ BSs, indexed by the set $\mathcal{B} = \{1, 2, \dots, B\}$. Each BS $b \in \mathcal{B}$ serves $K$ users within its coverage area. The $k$-th user served by BS $b$ is denoted as user $(b,k)$, where $k \in  \{1, 2, \dots, K\}$.

In terms of antenna configuration, each BS is equipped with $N$ FAs, which are connected to radio frequency (RF) chains via flexible cables. This setup allows the positions of the FAs to be adjusted dynamically within a specified two-dimensional (2D) transmit region, denoted by $\mathcal{C}_{t}$. Without loss of generality, the coordinate origin is placed at the lower-left corner of the square region, such that $\mathcal{C}_{t}=[0,A]\times[0,A]$. Let $\mathbf{t}_{b,n} = [x_{b,n}, y_{b,n}]^T \in \mathcal{C}_{t}$ represent the Cartesian coordinates of the $n$-th FA at BS $b$, for $n = 1, \dots, N$. The position matrix for all $N$ FAs at BS $b$ is denoted by $\tilde{\mathbf{t}}_b = [\mathbf{t}_{b,1}, \dots, \mathbf{t}_{b,N}] \in \mathbb{R}^{2 \times N}$. At the receiver side, each user is equipped with a single fixed-position antenna.

\subsection{Field-Response Based Channel Model}

 We adopt the far-field assumption, where the size of the transmit region is significantly smaller than the signal propagation distance. Therefore, for each multi-path component, the angle of departure (AoD), angle of arrival (AoA), and path gain magnitude remain invariant across the region. In contrast, the phase shifts vary dynamically according to the distinct spatial locations of the FAs \cite{10328751}.

 Let $L_{j,b,k}$ denote the number of propagation paths from the transmit region of BS $j$ to the receiver of user $(b,k)$.
For the $p$-th path ($p=1,\dots,L_{j,b,k}$), the AoD is characterized by the elevation angle $\theta_{j,b,k}^p \in [0, \pi]$ and the azimuth angle $\phi_{j,b,k}^p \in [0, \pi]$. Consequently, the phase difference of the signal propagation for the $p$-th transmit path between the $n$-th FA at position $\mathbf{t}_{j,n} = [x_{j,n}, y_{j,n}]^T$ and the origin of the transmit region is given by \cite{10243545}
\begin{equation}
\rho_{j,b,k}^p(\mathbf{t}_{j,n}) = x_{j,n} \sin\theta_{j,b,k}^p \cos\phi_{j,b,k}^p + y_{j,n} \cos\theta_{j,b,k}^p.
\end{equation}
Accordingly, the field response vector of the $n$-th FA at BS $j$, corresponding to the channel paths from BS $j$ to user $(b,k)$, is defined as
\begin{equation}
\mathbf{g}_{j,b,k}(\mathbf{t}_{j,n}) = \left[ e^{j\frac{2\pi}{l}\rho_{j,b,k}^1(\mathbf{t}_{j,n})}, \dots, e^{j\frac{2\pi}{l}\rho_{j,b,k}^{L_{j,b,k}}(\mathbf{t}_{j,n})} \right]^T,
\end{equation}
where $l$ is the carrier wavelength. By stacking the response vectors of all $N$ FAs, the field response matrix at BS $j$ is expressed as
\begin{equation}
\mathbf{G}_{j,b,k}(\tilde{\mathbf{t}}_j) = [\mathbf{g}_{j,b,k}(\mathbf{t}_{j,1}), \dots, \mathbf{g}_{j,b,k}(\mathbf{t}_{j,N})].
\end{equation}

Similarly, let $\mathbf{u}_{b,k} = [x_{b,k}, y_{b,k}]^T$, $\vartheta_{j,b,k}^p$, and $\varphi_{j,b,k}^p$ denote the fixed position of user $(b,k)$, the elevation angle, and the azimuth angle of the AoA for the $p$-th path, respectively. The phase difference at the receiver side relative to the origin of the user's coordinate system is calculated as \cite{10416896}
\begin{equation}
\nu_{j,b,k}^p(\mathbf{u}_{b,k}) = x_{b,k} \sin\vartheta_{j,b,k}^p \cos\varphi_{j,b,k}^p + y_{b,k} \cos\vartheta_{j,b,k}^p.
\end{equation}
The fixed receive response vector $\mathbf{r}_{b,k} \in \mathbb{C}^{L_{j,b,k} \times 1}$ at user $(b,k)$ is thus constructed as
\begin{equation}
\mathbf{r}_{b,k} = \left[ e^{j\frac{2\pi}{l}\nu_{j,b,k}^1(\mathbf{u}_{b,k})}, \dots, e^{j\frac{2\pi}{l}\nu_{j,b,k}^{L_{j,b,k}}(\mathbf{u}_{b,k})} \right]^T.
\end{equation}
Since the user antenna is fixed, the receiving response is constant. The physical channel vector $\mathbf{h}_{j,b,k}(\tilde{\mathbf{t}}_j) \in \mathbb{C}^{N \times 1}$ from BS $j$ to user $(b,k)$ is finally given by
\begin{equation}
\mathbf{h}_{j,b,k}^H(\tilde{\mathbf{t}}_j) = \mathbf{r}_{b,k}^{\text{H}} \mathbf{\Sigma}_{j,b,k} \mathbf{G}_{j,b,k}(\tilde{\mathbf{t}}_j),
\end{equation}
where $\mathbf{\Sigma}_{j,b,k} \in \mathbb{C}^{L_{j,b,k} \times L_{j,b,k}}$ is the diagonal path response matrix \cite{10243545} accounting for path losses and complex channel gains.

\subsection{Signal Transmission with Hardware Impairment Coefficients}
Let $a_{b,n}\in[a_{\min},1]$, where $a_{\min}>0$, denote the hardware impairment coefficient for the $n$-th FA at BS $b$. The positive lower bound represents the minimum operational radiation efficiency and yields a closed physical parameter set. In most conventional works, $a_{b,n}$ is simply regarded as 1. For mathematical tractability, we define the hardware impairment coefficient matrix $\mathbf{A}_b \in \mathbb{R}^{N \times N}$ for BS $b$ as
\begin{equation}
\mathbf{A}_b = \text{diag}\left( a_{b,1}, a_{b,2}, \dots, a_{b,N} \right).
\end{equation}
BS $b$ transmits a superposition of data streams to its $K$ served users. To distinguish the transmitted signal from the vectorized FA position, denote the baseband signal vector by $\mathbf{z}_b\in\mathbb{C}^{N\times1}$ \cite{10615837}, which is given by
\begin{equation}
\mathbf{z}_b = \sum_{k=1}^K \mathbf{w}_{b,k} s_{b,k},
\end{equation}
where $\mathbf{w}_{b,k} \in \mathbb{C}^{N \times 1}$ is the beamforming vector for user $(b,k)$, and $s_{b,k} \sim \mathcal{CN}(0,1)$ is the  transmitted symbol intended for the user $(b,k)$.
The actual radiated signal vector $\tilde{\mathbf{z}}_b$ is expressed as
\begin{equation}
\tilde{\mathbf{z}}_b = \mathbf{A}_b \mathbf{z}_b = \mathbf{A}_b \sum_{k=1}^K \mathbf{w}_{b,k} s_{b,k}.
\end{equation}

The received signal $y_{b,k}$ at user $(b,k)$ is a superposition of the signals transmitted from all BSs, propagated through their respective paths  \cite{10000992}. It can be expressed as
\begin{equation}
\begin{aligned}
y_{b,k} &= \sum_{j=1}^B \mathbf{h}_{j,b,k}^{\text{H}}(\tilde{\mathbf{t}}_j) \tilde{\mathbf{z}}_j + n_{b,k} \\
&= \mathbf{h}_{b,b,k}^{\text{H}}(\tilde{\mathbf{t}}_b) \mathbf{A}_b \mathbf{w}_{b,k} s_{b,k} + \sum_{i \neq k} \mathbf{h}_{b,b,k}^{\text{H}}(\tilde{\mathbf{t}}_b) \mathbf{A}_b \mathbf{w}_{b,i} s_{b,i} \\
&\quad + \sum_{j \neq b} \sum_{m=1}^K \mathbf{h}_{j,b,k}^{\text{H}}(\tilde{\mathbf{t}}_j) \mathbf{A}_j \mathbf{w}_{j,m} s_{j,m} + n_{b,k},
\end{aligned}
\end{equation}
where  $\mathbf{h}_{b,b,k}^{\text{H}}(\tilde{\mathbf{t}}_b) \mathbf{A}_b \mathbf{w}_{b,k} s_{b,k}$ is the desired signal,   $\sum_{i \neq k} \mathbf{h}_{b,b,k}^{\text{H}}(\tilde{\mathbf{t}}_b) \mathbf{A}_b \mathbf{w}_{b,i} s_{b,i}$ is the intra-cell interference,    $\sum_{j \neq b} \sum_{m=1}^K \mathbf{h}_{j,b,k}^{\text{H}}(\tilde{\mathbf{t}}_j) \mathbf{A}_j \mathbf{w}_{j,m} s_{j,m}$ is the inter-cell
interference, and   $n_{b,k} \sim \mathcal{CN}(0, \sigma^2)$ denotes the additive white Gaussian noise (AWGN) at the receiver.

Consequently, the signal-to-interference-plus-noise ratio (SINR) for user $(b,k)$ is derived as
\begin{equation}
\gamma_{b,k} = \frac{\left| \mathbf{h}_{b,b,k}^{\text{H}}(\tilde{\mathbf{t}}_b) \mathbf{A}_b \mathbf{w}_{b,k} \right|^2}{ I_{b,k}^{\text{intra}} + I_{b,k}^{\text{inter}} + \sigma^2 },
\end{equation}
where $I_{b,k}^{\text{intra}} = \sum_{i \neq k} \left| \mathbf{h}_{b,b,k}^H(\tilde{\mathbf{t}}_b) \mathbf{A}_b \mathbf{w}_{b,i} \right|^2$ represents the intra-cell interference and $I_{b,k}^{\text{inter}} = \sum_{j \neq b} \sum_{m=1}^K \left| \mathbf{h}_{j,b,k}^H(\tilde{\mathbf{t}}_j) \mathbf{A}_j \mathbf{w}_{j,m} \right|^2$ represents the inter-cell interference.

\subsection{Forward Problem Formulation}
The goal of the forward problem is to maximize the SE of the  FAS by jointly optimizing the FA position $\tilde{\mathbf{t}}_b$ and the transmit beamforming  vector $\mathbf{w}_{b,k}$. The forward optimization problem is formulated as follows

\begin{subequations}  \label{OP1}
\begin{align}
 \max_{\{\tilde{\mathbf{t}}_b\}, \{\mathbf{w}_{b,k}\}} \quad & \sum_{b=1}^B \sum_{k=1}^K \log_2 \left( 1 + \gamma_{b,k} \right) \tag{\theequation} \\
\text{s.t.} \quad & \mathbf{t}_{b,n} \in \mathcal{C}_{t}, \quad \forall b \in \mathcal{B}, \forall n \in \{1,\dots,N\}, \label{eq:constraint_region} \\
& \|\mathbf{t}_{b,n} - \mathbf{t}_{b,l}\|_2 \ge D, \quad \forall b \in \mathcal{B}, \forall n \neq l, \label{eq:constraint_dist} \\
& \sum_{k=1}^K \|\mathbf{w}_{b,k}\|_2^2 \le P_{max}, \quad \forall b \in \mathcal{B}, \label{eq:constraint_power}
\end{align}
\end{subequations}
where $P_{max}$ is the  transmit power of the BS and  $D$ is the minimum distance between FAs. Constraint (\ref{OP1}a) ensures that all FAs are located within the feasible transmit region $\mathcal{C}_t$. Constraint (\ref{OP1}b) enforces a minimum distance $D$ between  FAs at the same BS to avoid physical coupling.   The transmit power constraint of
each BS is given in (\ref{OP1}c).
Problem (\ref{OP1}) is non-convex because the interference terms and separation constraints are coupled. Next,
	we introduce an efficient scheme to solve problem (\ref{OP1}).

\section{ Inverse Optimization Model}

\subsection{Consistency-Priced Interference Representation}
In problem (\ref{OP1}), the interference generated by BS $j$ toward user $(b,k)$ is
\begin{equation}
I_{j \to b,k}(\tilde{\mathbf{t}}_j,\mathbf{W}_j)=\sum_{m=1}^{K}\left|\mathbf{h}_{j,b,k}^{H}(\tilde{\mathbf{t}}_j)\mathbf{A}_j\mathbf{w}_{j,m}\right|^{2}.
\end{equation}
Following \cite{1661442}, we introduce an aggregate variable $q_{b,k}$ satisfying
\begin{equation}
q_{b,k}\geq\sum_{j\neq b}I_{j\to b,k}(\tilde{\mathbf{t}}_j,\mathbf{W}_j).
\label{eq:aggregate_interference}
\end{equation}
To verify tightness, let $S_{b,k}$ and $J_{b,k}$ denote the desired-signal power and the sum of intra-cell interference and noise, respectively. The achievable rate can then be written as
\begin{equation}
R_{b,k}=\log_2\left(1+\frac{S_{b,k}}{J_{b,k}+q_{b,k}}\right).
\end{equation}
For fixed physical variables,
\begin{equation}
\frac{\partial R_{b,k}}{\partial q_{b,k}}
=-\frac{S_{b,k}}{\ln 2\,(J_{b,k}+q_{b,k})(J_{b,k}+q_{b,k}+S_{b,k})}<0.
\label{eq:rate_q_derivative}
\end{equation}
Hence, any strict slack in (\ref{eq:aggregate_interference}) can be reduced to increase the objective, and equality holds at an optimum. Let $\lambda_{b,k}\geq0$ be the associated consistency price. After dualizing (\ref{eq:aggregate_interference}), the terms involving the local variables of BS $b$ can be expressed as
\begin{equation}
\mathcal L_b=\sum_{k}R_{b,k}+\sum_k\lambda_{b,k}q_{b,k}-\sum_{j\neq b}\sum_k\lambda_{j,k}I_{b\to j,k}.
\label{eq:local_lagrangian}
\end{equation}
The first two terms in (\ref{eq:local_lagrangian}) reward the served users, whereas the last term charges BS $b$ for the interference imposed on users in the other cells. At iteration $r$, each BS maximizes a concave successive convex approximation (SCA) surrogate of (\ref{eq:local_lagrangian}) over its local position and power constraints. The BSs then exchange the scalar interference contributions, and the prices are updated according to
\begin{equation}
\lambda_{b,k}^{(r+1)}=\left[\lambda_{b,k}^{(r)}-\alpha_r\left(q_{b,k}^{(r)}-\sum_{j\neq b}I_{j\to b,k}^{(r)}\right)\right]^+,
\label{eq:price_update}
\end{equation}
where $\alpha_r>0$ is a diminishing step size and $[x]^+=\max\{x,0\}$. The corresponding price decreases when the local copy exceeds the physical aggregate interference, whereas it increases otherwise to penalize the BSs responsible for the mismatch. Thus, the distributed routine enforces physical interference consistency at a stationary point of the SCA iterations and does not optimize an unconstrained interference estimate.

In the inverse stage, $q_{b,k}$ is not a decision variable. Instead, $\hat q_{b,k}$ is measured through coordinated reference-signal or BS-muting intervals under the recorded positions and beamformers. This avoids evaluating (\ref{eq:aggregate_interference}) using the unknown hardware coefficients. The tuples $\{\hat{\mathbf t}_b,\hat{\mathbf W}_b,\hat{\mathbf q}_b\}$ and terminal prices are recorded within the same channel-coherence interval.

\subsection{Position Model Conditioned on Observed Beamforming}
For a multi-user interference channel, we retain the observed beamforming matrices $\hat{\mathbf W}=\{\hat{\mathbf W}_j\}_{j=1}^{B}$ \cite{10196368}. Define $\hat R_{b,k}(\tilde{\mathbf t}_b;\mathbf A_b)=R_{b,k}(\tilde{\mathbf t}_b,\hat{\mathbf W},\hat{\mathbf q};\mathbf A_b)$. With $\hat{\mathbf q}$ and the converged prices fixed, constants independent of $\tilde{\mathbf t}_b$ can be omitted, yielding
\begin{equation}
\begin{aligned}
\bar{\mathcal L}_b(\tilde{\mathbf t}_b;\mathbf A_b)
=&\sum_k\hat R_{b,k}(\tilde{\mathbf t}_b;\mathbf A_b)\\
&-\sum_{j\neq b}\sum_k\hat\lambda_{j,k}I_{b\to j,k}(\tilde{\mathbf t}_b,\hat{\mathbf W}_b).
\end{aligned}
\label{eq:conditional_position}
\end{equation}
This conditional model preserves intra-cell interference and the priced inter-cell externality. The observed $\hat{\mathbf W}$ can be produced by WMMSE and is held fixed when differentiating with respect to the FA positions. Thus, the inverse model explains the observed position update without asserting a closed-form or globally optimal beamformer for (\ref{OP1}).

\subsection{Trust-Region Linear Forward Model}
Let $\mathbf x_b=\operatorname{vec}(\tilde{\mathbf t}_b)$ and $\hat{\mathbf x}_b=\operatorname{vec}(\hat{\mathbf t}_b)$. The cost vector of the local first-order model is
\begin{equation}
\mathbf c(\mathbf A_b)=-\nabla_{\mathbf x_b}\bar{\mathcal L}_b(\mathbf x_b;\mathbf A_b)\big|_{\mathbf x_b=\hat{\mathbf x}_b}.
\label{eq:cost_gradient}
\end{equation}
For each antenna pair $n<l$, define $\hat{\mathbf d}_{nl}=\hat{\mathbf t}_{b,n}-\hat{\mathbf t}_{b,l}$ and $\mathbf u_{nl}=\hat{\mathbf d}_{nl}/\|\hat{\mathbf d}_{nl}\|_2$. Convexity of the Euclidean norm gives the global affine lower bound
\begin{equation}
\|\mathbf t_{b,n}-\mathbf t_{b,l}\|_2
\geq\mathbf u_{nl}^{T}(\mathbf t_{b,n}-\mathbf t_{b,l}),
\label{eq:distance_lower_bound}
\end{equation}
so imposing $\mathbf u_{nl}^{T}(\mathbf t_{b,n}-\mathbf t_{b,l})\geq D$ is an inner approximation that guarantees the original separation constraint. The trust region $\|\mathbf x_b-\hat{\mathbf x}_b\|_\infty\leq\Delta$ restricts every coordinate to the neighborhood in which the objective linearization is informative.

For completeness, let $\mathbf S_n\in\mathbb R^{2\times2N}$ select the $n$-th antenna position, i.e., $\mathbf t_{b,n}=\mathbf S_n\mathbf x_b$, and stack the rows $\mathbf u_{nl}^{T}(\mathbf S_n-\mathbf S_l)$, $n<l$, into $\mathbf D_{\rm sep}$. Since $\mathcal C_t=[0,A]^2$, all constraints can be written explicitly as
\begin{equation}
\begin{bmatrix}
\mathbf I_{2N}\\-\mathbf I_{2N}\\\mathbf D_{\rm sep}\\
\mathbf I_{2N}\\-\mathbf I_{2N}
\end{bmatrix}\mathbf x_b\geq
\begin{bmatrix}
\mathbf0\\-A\mathbf1\\D\mathbf1\\
\hat{\mathbf x}_b-\Delta\mathbf1\\-\hat{\mathbf x}_b-\Delta\mathbf1
\end{bmatrix}.
\label{eq:explicit_De}
\end{equation}
The five block rows in (\ref{eq:explicit_De}) represent the two region bounds, pairwise separation, and the two trust-region bounds. Here, $\mathbf I_{2N}$ has dimension $2N$, whereas $\mathbf D_{\rm sep}$ has $N(N-1)/2$ rows. The local linear forward problem is
\begin{equation}
\min_{\mathbf x_b}\ \mathbf c(\mathbf A_b)^T\mathbf x_b\quad\text{s.t.}\quad\mathbf D\mathbf x_b\geq\mathbf e,
\label{eq:compact_forward}
\end{equation}
where
\begin{equation*}
\mathbf D=\begin{bmatrix}
\mathbf I_{2N}\\-\mathbf I_{2N}\\\mathbf D_{\rm sep}\\
\mathbf I_{2N}\\-\mathbf I_{2N}
\end{bmatrix},\qquad
\mathbf e=\begin{bmatrix}
\mathbf0\\-A\mathbf1\\D\mathbf1\\
\hat{\mathbf x}_b-\Delta\mathbf1\\-\hat{\mathbf x}_b-\Delta\mathbf1
\end{bmatrix}.
\end{equation*}
 The trust-region radius $\Delta$ can be decreased whenever the first-order model poorly predicts the actual objective change and enlarged after successful steps.

\subsection{Inverse Optimization With Inexact Observations}
The dual of (\ref{eq:compact_forward}) is $\max_{\boldsymbol\mu\geq\mathbf0}\mathbf e^T\boldsymbol\mu$ subject to $\mathbf D^T\boldsymbol\mu=\mathbf c(\mathbf A_b)$. To allow a feasible observed position to be generated by an inexact or heuristic forward algorithm, we introduce a nonnegative duality-gap tolerance $\epsilon$ and formulate
\begin{subequations}\label{IOPF}
\begin{align}
\min_{\mathbf A_b,\boldsymbol\mu,\epsilon}\quad&\frac{1}{2}\|\mathbf c(\mathbf A_b)-\hat{\mathbf c}_{prior}\|_2^2+\rho\epsilon \tag{\theequation}\\
\text{s.t.}\quad&\mathbf D\hat{\mathbf x}_b\geq\mathbf e,\label{eq:iop_primal_feas}\\
&\mathbf D^T\boldsymbol\mu=\mathbf c(\mathbf A_b),\label{eq:iop_stat}\\
&0\leq\mathbf c(\mathbf A_b)^T\hat{\mathbf x}_b-\mathbf e^T\boldsymbol\mu\leq\epsilon,\label{eq:iop_gap}\\
&\boldsymbol\mu\geq\mathbf0,\quad\epsilon\geq0,\label{eq:iop_mu_pos}\\
&a_{\min}\mathbf1\leq\operatorname{diag}(\mathbf A_b)\leq\mathbf1,\quad a_{\min}>0.\label{eq:iop_physical}
\end{align}
\end{subequations}
Here, $\hat{\mathbf c}_{prior}$ is obtained from calibration data, and $\rho>0$ controls the fit--optimality tradeoff. By weak duality, the gap in (\ref{eq:iop_gap}) is nonnegative. When $\epsilon=0$, it reduces to strong duality and is equivalent to complementary slackness. A positive $\epsilon$ quantifies the admissible suboptimality of a heuristic position, while the objective penalizes a large inconsistency. The closed bounds in (\ref{eq:iop_physical}) exclude nonphysical or numerically singular coefficients.

Problem (\ref{IOPF}) is a smooth nonlinear program when $\mathbf c(\mathbf A_b)$ is continuously differentiable. Let $\boldsymbol\zeta=[\operatorname{diag}(\mathbf A_b)^T,\boldsymbol\mu^T,\epsilon]^T$, with  its equality and inequality constraints represented by $\mathbf h(\boldsymbol\zeta)=\mathbf0$ and $\mathbf g(\boldsymbol\zeta)\geq\mathbf0$, respectively. At SQP iteration $r$, the search direction is obtained from
\begin{equation}
\begin{aligned}
\min_{\mathbf d}\quad&\nabla f(\boldsymbol\zeta^{(r)})^T\mathbf d
+\tfrac12\mathbf d^T\mathbf B_r\mathbf d\\
\text{s.t.}\quad&\mathbf h(\boldsymbol\zeta^{(r)})
+\nabla\mathbf h(\boldsymbol\zeta^{(r)})^T\mathbf d=\mathbf0,\\
&\mathbf g(\boldsymbol\zeta^{(r)})
+\nabla\mathbf g(\boldsymbol\zeta^{(r)})^T\mathbf d\geq\mathbf0,
\end{aligned}
\label{eq:sqp_subproblem}
\end{equation}
where $f$ is the objective of (\ref{IOPF}) and $\mathbf B_r$ is a positive-definite Broyden-Fletcher-Goldfarb-Shanno (BFGS) approximation of the  Hessian of the Lagrangian. A merit-function backtracking line search selects the step length, after which the BFGS matrix and constraint multipliers are updated. The procedure terminates when the primal feasibility, stationarity, and step-size residuals are below prescribed tolerances.

\subsection{Convergence Analysis}
\subsubsection{Global Convergence} Assume that the iterates remain in a bounded level set, the linearized constraints are feasible, the functions in (\ref{IOPF}) are continuously differentiable, and the linear independence constraint qualification holds at every accumulation point. With a sufficiently large merit-function penalty and uniformly positive-definite BFGS matrices, standard line-search SQP results ensure that every accumulation point is a KKT point of (\ref{IOPF}). 

\subsubsection{Local Convergence Rate} If a KKT point additionally satisfies strict complementarity and the strong second-order sufficient condition, and the BFGS updates satisfy the Dennis--Mor\'{e} condition, the SQP iterates converge locally at a superlinear rate.

\section{Simulation Results And Analysis}
For our simulations, we consider a  multi-BS multi-user communication system consisting of 4 BSs. Each BS is equipped with 4 FAs and serves 4 users unless the number of users is varied explicitly. The true hardware coefficients are drawn independently from $\mathcal U[a_{\min},0.95]$.  The elements of the diagonal path response matrix are modeled as
$\mathbf{\Sigma}[1,1]\sim \mathcal{CN}\left(0, \frac{\kappa}{\kappa+1}\right)$ and $\mathbf{\Sigma}[p,p]\sim \mathcal{CN}\left(0, \frac{1}{(\kappa + 1)(L_{j,b,k} - 1)}\right),\text{for }  p = 2, 3, \dots, L_{j,b,k}$, where $\kappa$ is the ratio of line-of-sight to non-line-of-sight  signal power.
Other parameters are listed in Table I.

\begin{table}[t]
	\renewcommand{\arraystretch}{1.18}
	\centering
	\caption{Simulation Parameters}
	\label{table2}
	\begin{tabular}{|c|c|c|}
		\hline
		\textbf{Parameters}   & \textbf{Notation}    & \textbf{Values}     \\ \hline
	Transmit powers of the BS  &$P_{\max}$ & 20 and 30 dBm  \\ \hline
		Noise power    & $\sigma^2$  & -90 dBm              \\ \hline
		Propagation paths from the BS to users    &  $L_{j,b,k}$   & 4        \\ \hline
	Minimum distance between FAs    & $D$   & $l/2$          \\ \hline
    Region size     & $A$   & $4l$ \\ \hline
	Hardware coefficient lower bound & $a_{\min}$ & $0.5$ \\ \hline
	Trust-region radius & $\Delta$ & $0.25l$ \\ \hline
	Duality-gap weight & $\rho$ & $1$ \\ \hline
Stopping tolerance & $\tau$ & $10^{-6}$ \\ \hline
	\end{tabular}
\end{table}

For comparison purposes, we consider the following three baselines:

\begin{itemize}
\item
KKT Residual Minimization (KKT-RM): This inverse-optimization benchmark uses the same observed $\hat{\mathbf W}$ and $\hat{\mathbf q}$, local model $(\mathbf c,\mathbf D,\mathbf e)$, and coefficient bounds as the proposed scheme, but incorporates the stationarity and duality-gap conditions as regularized residual penalties rather than enforcing them as explicit constraints.

\item
Ideal-hardware-assumption scheme: This scheme sets $\mathbf A_b=\mathbf I$ when designing the FA positions and beamformers, while the resulting decisions are evaluated using the true impaired hardware coefficients.

\item
Perfect coefficient scheme: This scheme uses the ground-truth hardware coefficients for position and beamforming design and evaluates the resulting decisions under the same true coefficients, thereby providing an oracle performance reference.

\end{itemize}

\begin{figure}[!t]
	\centering
	\includegraphics[width=7cm]{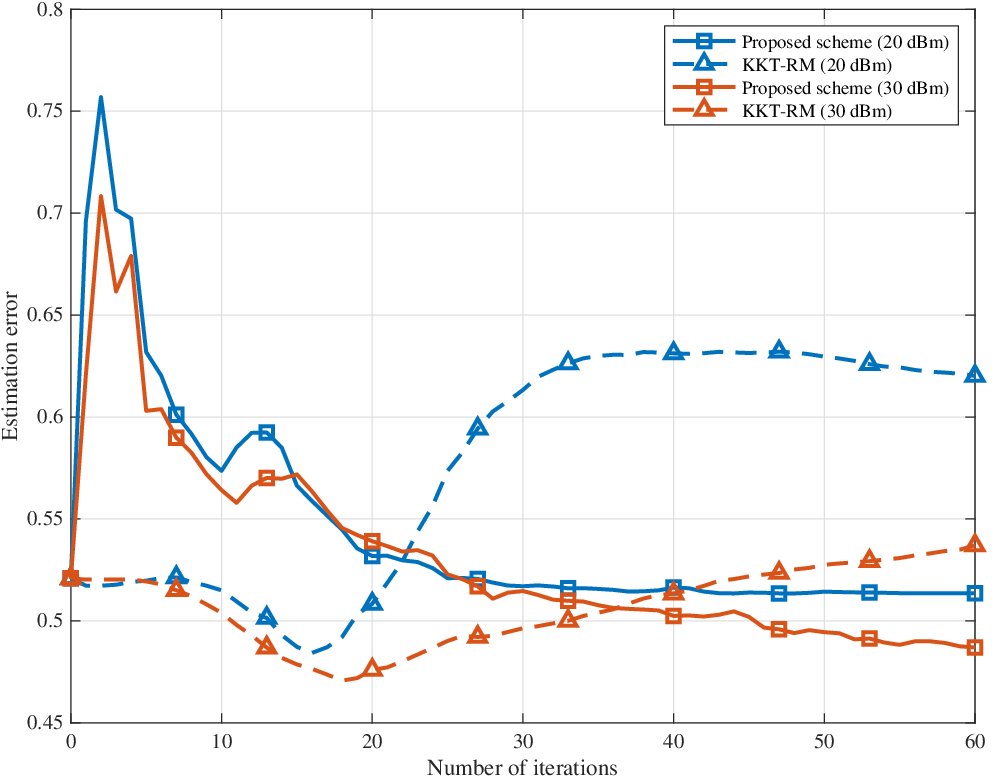}\\
	\caption{Estimation error versus the number of iterations.}\label{convergency}
\end{figure}

Fig. \ref{convergency} shows the estimation error versus the number of iterations, where the error is defined as the Euclidean norm between the estimated and ground-truth hardware-coefficient vectors across all BSs. Although transient increases occur during the early iterations, the proposed scheme subsequently decreases and stabilizes, whereas KKT-RM decreases initially and then exhibits a pronounced error rebound. At iteration 60, the proposed scheme attains errors of approximately $0.514$ and $0.487$ at 20 and 30 dBm, respectively, compared with $0.620$ and $0.537$ for KKT-RM. These values correspond to final-error reductions of about 17.2\% and 9.3\%, respectively. The improvement results from enforcing stationarity and the admissible duality gap directly in the SQP subproblems, while KKT-RM trades these conditions against the data-fitting term through finite residual penalties.

\begin{figure}[!t]
	\centering
	\includegraphics[width=7.3cm]{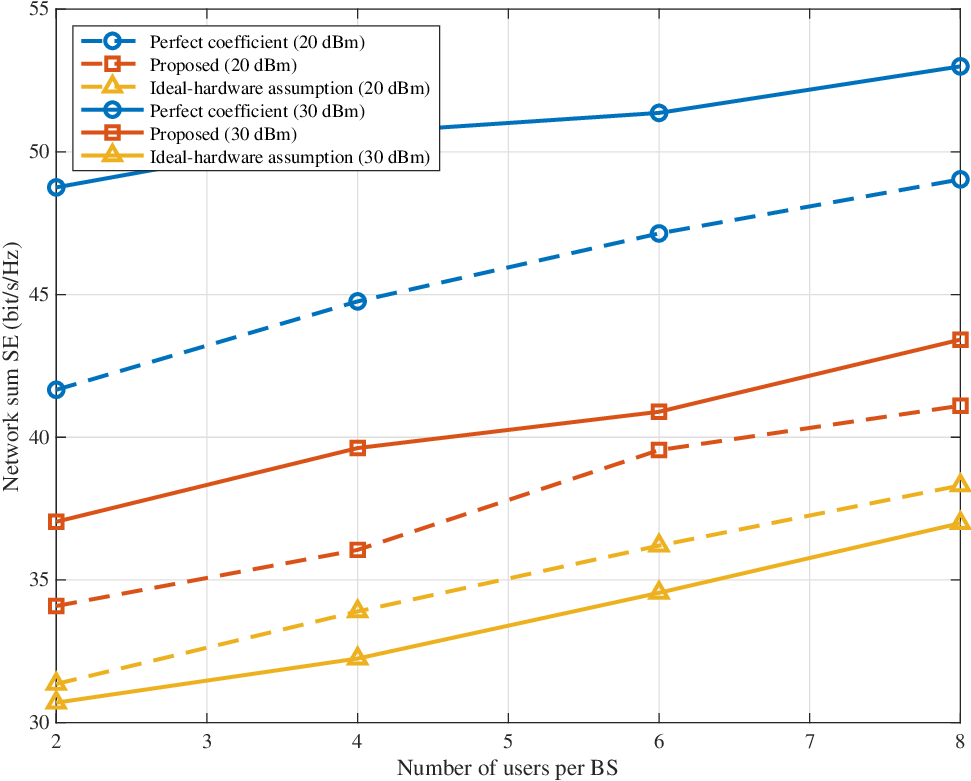}\\
	\caption{Network sum SE versus the number of users per BS.}\label{untitled_curve}
\end{figure}

Fig. \ref{untitled_curve} shows the network sum SE versus the number of users per BS. For all schemes, the  SE is evaluated using the true impaired channels. The schemes differ only in the hardware impairment coefficients used for position and beamforming design. The sum SE increases with the number of users for both transmit-power levels, and the perfect coefficient scheme provides the highest performance. The proposed scheme consistently outperforms the Ideal-hardware-assumption scheme. Specifically, its gains range from 6.37\% to 9.24\% at 20 dBm and from 17.38\% to 22.86\% at 30 dBm, with the largest gain obtained for four users per BS. The larger gain at 30 dBm indicates that coefficient mismatch becomes more detrimental as the interference generated by the mismatched design increases. This effect also explains why the 30 dBm Ideal-hardware-assumption curve is slightly below its 20 dBm counterpart, while the proposed and perfect coefficient schemes benefit from the additional transmit power.

\section{Conclusions}
In this paper, we have developed an inverse optimization framework for estimating hardware impairment coefficients of FAs. Aggregate interference variables and consistency prices characterize the inter-cell coupling in the distributed forward routine, while inverse estimation uses the receiver-side interference measurements associated with the observed network decisions. By conditioning on the observed beamforming matrices, we avoid imposing a closed-form beamformer in the multi-user interference channel. A trust-region linear position model and a duality-gap relaxation accommodate locally optimized or heuristic position observations. The resulting nonlinear inverse problem is solved by SQP under standard regularity conditions.  Simulation results show that the proposed scheme achieves lower estimation errors than KKT-RM and improves the  SE by up to 22.86\% over the Ideal-hardware-assumption scheme.

\bibliographystyle{IEEEtran}
\bibliography{IEEEabrv,MMM}

\begin{thebibliography}{10}
\providecommand{\url}[1]{#1}
\csname url@samestyle\endcsname
\providecommand{\newblock}{\relax}
\providecommand{\bibinfo}[2]{#2}
\providecommand{\BIBentrySTDinterwordspacing}{\spaceskip=0pt\relax}
\providecommand{\BIBentryALTinterwordstretchfactor}{4}
\providecommand{\BIBentryALTinterwordspacing}{\spaceskip=\fontdimen2\font plus
\BIBentryALTinterwordstretchfactor\fontdimen3\font minus
  \fontdimen4\font\relax}
\providecommand{\BIBforeignlanguage}[2]{{%
\expandafter\ifx\csname l@#1\endcsname\relax
\typeout{** WARNING: IEEEtran.bst: No hyphenation pattern has been}%
\typeout{** loaded for the language `#1'. Using the pattern for}%
\typeout{** the default language instead.}%
\else
\language=\csname l@#1\endcsname
\fi
#2}}
\providecommand{\BIBdecl}{\relax}
\BIBdecl

\bibitem{9650760}
K.-K. Wong and K.-F. Tong, ``Fluid antenna multiple access,'' \emph{IEEE Trans.
  Wireless Commun.}, vol.~21, no.~7, pp. 4801--4815, Jul. 2022.

\bibitem{10753482}
W.~K. New, K.-K. Wong, H.~Xu, C.~Wang, F.~R. Ghadi, J.~Zhang, J.~Rao, R.~Murch,
  P.~Ramírez-Espinosa, D.~Morales-Jimenez, C.-B. Chae, and K.-F. Tong, ``A
  tutorial on fluid antenna system for 6{G} networks: Encompassing
  communication theory, optimization methods and hardware designs,'' \emph{IEEE
  Commun. Surveys Tuts.}, vol.~27, no.~4, pp. 2325--2377, Aug. 2025.

\bibitem{11098630}
J.~Huangfu, Z.~Song, T.~Hou, A.~Li, Y.~Liu, A.~Nallanathan, and K.-K. Wong,
  ``Performance analysis of fluid antenna system under spatially-correlated
  rician fading channels,'' \emph{IEEE Trans. Wireless Commun.}, vol.~25, pp.
  1394--1407, Jul. 2026.

\bibitem{10130117}
W.~K. New, K.-K. Wong, H.~Xu, K.-F. Tong, and C.-B. Chae, ``Fluid antenna
  system: New insights on outage probability and diversity gain,'' \emph{IEEE
  Trans. Wireless Commun.}, vol.~23, no.~1, pp. 128--140, Jan. 2024.

\bibitem{xu2023capacity}
H.~Xu, K.-K. Wong, W.~K. New, F.~R. Ghadi, G.~Zhou, R.~Murch, C.-B. Chae,
  Y.~Zhu, and S.~Jin, ``Capacity maximization for {FAS}-assisted multiple
  access channels,'' \emph{IEEE Trans. Commun.}, vol.~73, no.~7, pp.
  4713--4731, Jul. 2025.

\bibitem{8581475}
S.~Javed, O.~Amin, S.~S. Ikki, and M.-S. Alouini, ``Multiple antenna systems
  with hardware impairments: New performance limits,'' \emph{IEEE Trans. Veh.
  Technol.}, vol.~68, no.~2, pp. 1593--1606, Feb. 2019.

\bibitem{11224643}
J.~Yao, T.~Wu, L.~Zhou, M.~Jin, C.~Pan, M.~Elkashlan, F.~Adachi, G.~K.
  Karagiannidis, N.~Al-Dhahir, and C.~Yuen, ``Rethinking hardware impairments
  in multi-user systems: Can {FAS} make a difference?'' \emph{IEEE Trans. Mob.
  Comput.}, vol.~25, no.~4, pp. 4965--4978, Apr. 2026.

\bibitem{chan2020inverse}
T.~C. Chan and N.~Kaw, ``Inverse optimization for the recovery of constraint
  parameters,'' \emph{Eur. J. Oper. Res.}, vol. 282, no.~2, pp. 415--427, Apr.
  2020.

\bibitem{9336679}
S.~A. Akhtar, A.~S. Kolarijani, and P.~M. Esfahani, ``Learning for control: An
  inverse optimization approach,'' \emph{IEEE Control Syst. Lett.}, vol.~6, pp.
  187--192, Jan. 2021.

\bibitem{chan2025inverse}
T.~C. Chan, R.~Mahmood, and I.~Y. Zhu, ``Inverse optimization: Theory and
  applications,'' \emph{Oper. Res.}, vol.~73, no.~2, pp. 1046--1074, Apr. 2025.

\bibitem{ahuja2001inverse}
R.~K. Ahuja and J.~B. Orlin, ``Inverse optimization,'' \emph{Oper. Res.},
  vol.~49, no.~5, pp. 771--783, Oct. 2001.

\bibitem{mohajerin2018data}
P.~Mohajerin~Esfahani, S.~Shafieezadeh-Abadeh, G.~A. Hanasusanto, and D.~Kuhn,
  ``Data-driven inverse optimization with imperfect information,'' \emph{Math.
  Program.}, vol. 167, no.~1, pp. 191--234, Dec. 2018.

\bibitem{7859377}
J.~Saez-Gallego and J.~M. Morales, ``Short-term forecasting of price-responsive
  loads using inverse optimization,'' \emph{IEEE Trans. Smart Grid}, vol.~9,
  no.~5, pp. 4805--4814, Sep. 2018.

\bibitem{10328751}
Y.~Ye, L.~You, J.~Wang, H.~Xu, K.-K. Wong, and X.~Gao, ``Fluid antenna-assisted
  {MIMO} transmission exploiting statistical {CSI},'' \emph{IEEE Commun.
  Lett.}, vol.~28, no.~1, pp. 223--227, Jan. 2024.

\bibitem{10243545}
W.~Ma, L.~Zhu, and R.~Zhang, ``{MIMO} capacity characterization for movable
  antenna systems,'' \emph{IEEE Trans. Wireless Commun.}, vol.~23, no.~4, pp.
  3392--3407, Apr. 2024.

\bibitem{10416896}
H.~Qin, W.~Chen, Z.~Li, Q.~Wu, N.~Cheng, and F.~Chen, ``Antenna positioning and
  beamforming design for fluid antenna-assisted multi-user downlink
  communications,'' \emph{IEEE Wireless Commun. Lett.}, vol.~13, no.~4, pp.
  1073--1077, Apr. 2024.

\bibitem{10615837}
S.~Zhang, M.~Chen, W.~Zhang, Z.~Yang, D.~Liu, Z.~Zhang, and K.-K. Wong,
  ``Performance optimization for multicast mm{W}ave {MIMO} networks with mobile
  users,'' in \emph{Proc. IEEE Int. Conf. Commun. Workshops (ICC Workshops)},
  Denver, CO, USA, Aug. 2024, pp. 846--851.

\bibitem{10000992}
S.~Zhang, Z.~Yang, M.~Chen, D.~Liu, K.-K. Wong, and H.~V. Poor, ``Performance
  optimization for intelligent reflecting surface assisted multicast {MIMO}
  networks,'' in \emph{Proc. IEEE Glob. Commun. Conf. (GLOBECOM),}, Rio de
  Janeiro, Brazil, Dec. 2022, pp. 5838--5843.

\bibitem{1661442}
C.~W. Tan, D.~Palomar, and M.~Chiang, ``Distributed optimization of coupled
  systems with applications to network utility maximization,'' in \emph{Proc.
  IEEE Int. Conf. Acoust. Speech Signal Process (ICASSP)}, Toulouse, France,
  Jul. 2006, pp. 1--4.

\bibitem{10196368}
S.~Zhang, Z.~Yang, M.~Chen, D.~Liu, K.-K. Wong, and H.~V. Poor, ``Beamforming
  design for the performance optimization of intelligent reflecting surface
  assisted multicast {MIMO} networks,'' \emph{IEEE Trans. Wireless Commun.},
  vol.~23, no.~3, pp. 2325--2339, Mar. 2024.

\end{thebibliography}



\end{document}